\documentclass[11pt]{article}
\usepackage{amsfonts, amsopn, amsthm, amsmath}
\advance\voffset-2cm
\advance\hoffset-1.8cm
\advance\textheight4.9cm
\advance\textwidth2.9cm

\def\B{{\mathbb B}}
\def\C{{\mathbb C}}
\def\E{{E}}

\def\LO{{\cal O}}

\def\Q{{\mathbb Q}}

\def\SL{{\rm SL}(2,\Z)}
\def\Z{{\mathbb Z}}

\def\e{{\operatorname e}}
\def\w{{\cal W}}

\newtheorem*{conj}{Conjecture}

\begin{document}
\begin{flushright}
DAMTP-96-63 \\
hep-th/9607165
\end{flushright}

\begin{center}
{{\Large\sc Product expansions of conformal characters}}

\vspace{0.2cm}

{\large{Wolfgang Eholzer\footnote[1]{%
Supported by the EPSRC, partial support from 
PPARC and from EPSRC, grant GR/J73322, is also acknowledged.}
and Nils-Peter Skoruppa}}

\vspace{0.2cm}

{\small Department of Applied Mathematics and Theoretical Physics,\\
University of Cambridge, Silver Street,\\
Cambridge, CB3 9EW, U.K.\\
W.Eholzer@damtp.cam.ac.uk}

\vspace{0.2cm}

{and \\ }

\vspace{0.2cm}

{\small Universit\'e Bordeaux I,\\
U.F.R.\ de Math\'ematiques et Informatique,\\
351, Cours de la Lib\'eration,\\
33405 Talence, France\\
skoruppa@math.u-bordeaux.fr}

\end{center}

{\bf Abstract.}
We describe several infinite series of 
rational conformal field theories whose 
conformal characters are modular units, i.e.\  
which are modular functions having no zeros or 
poles in the upper complex half plane, and which
thus possess simple product expansions.
We conjecture that certain infinite series of
rational models of Casimir $\w$-algebras always have this property.
Furthermore, we describe an algorithm which can be used to 
prove whether a modular function is a modular unit or not.

\section{Introduction}

It is a well-known phenomenon that occasionally
the conformal characters of rational conformal field theories (RCFTs) in 
two dimensions have simple product expansions.
In this letter we make a first attempt to investigate this
more systematically. In particular, we shall explain
why such product expansions, if they exist at all, are
always of a certain simple shape.

The main observation for explaining the existence of such product 
expansions is that in many cases the conformal characters of 
RCFTs in two dimensions are 
not only modular functions, but even modular units, and that
modular units with integral Fourier coefficients
always have product expansions of a very simple type.

A modular unit is a modular function on some congruence 
subgroup of the modular group which has no poles or zeros in 
the upper complex half plane. The modular units obviously form
a group under multiplication. This group is 
generated by the so-called Siegel units and square roots
of certain products of these. The Siegel units are
classical elementary functions, which have 
nice product expansions.

It is believed that the conformal 
characters of rational models are modular functions on some 
congruence subgroup. If this were true, then every conformal 
character having no zeros or poles in the upper half plane could 
be expressed as a product of integral powers of Siegel units 
(or square roots of these).

In the context of conformal characters however, it is more natural
to look at the subgroup $\E$ of modular units whose Fourier expansions
are of the form $q^s(1+q\,\Z[\![q]\!])$ with rational numbers $s$. Here
$q^s(z)=\e^{2\pi i sz}$, where $z$ is the variable in the upper half plane.
A modular unit with Fourier expansion of the form $q^s \Z[\![q]\!]$ ($s\in\Q$)
is always an element of the group $\Q^\ast\cdot\E$ \cite{ES3}.
Conformal characters which are modular units are thus, 
up to multiplication by a positive integer, 
elements of the semigroup $\E^+$ of all units in $\E$ which have non-negative
Fourier coefficients.

Special instances of $\E^+$ are the functions
$$
[r]_l=
q^{-l\B_2\left(\frac rl\right)/2}
\,
\prod_{
\begin{subarray}{c}
n\equiv +r\bmod l\\
n>0
\end{subarray}
}
\left(1-q^n\right)^{-1}
\prod_{
\begin{subarray}{c}
n\equiv -r\bmod l\\
n>0
\end{subarray}
}
\left(1-q^n\right)^{-1},
$$
where $l\ge 1$ and $r$ are integers with $r$ not divisible
by $l$, and where
$\B_2(x)=y^2-y+1/6$ with $y=x-\lfloor x\rfloor$.
Note that for even $l$ and $r=l/2$ both products on the right hand side
are equal.

In \cite[Theorem 1]{ES3} it was proved that the group $\E$ is generated 
by the functions $[r]_l$.
In other words, a conformal character $\chi$
is a modular unit, if and only if $\chi$ can be written
as a product of integral powers of the functions $[r]_l$ (times a constant).
This explains the simple shape of the product expansions
of conformal characters which are modular units.
However, the question why they are modular units at all
remains open.

A closer look shows that in many cases even all conformal characters
of a RCFT are modular units. Even more, in all but one of such
examples which are known to us, the conformal characters can be 
written as products of positive integral powers of functions $[r]_l$.
Let $\E_{\ast}$ be the semigroup of all such products, i.e.\ the semigroup
generated by the functions $[r]_l$.
As already mentioned, we do not know what it means physically that a 
conformal character is a modular unit. However,
in view of the fact that most examples for this phenomenon
already lie in $\E_{\ast}$, one 
might speculate that this phenomenon is related to 
the existence of distinguished bases in the corresponding 
representation spaces (cf.\ end of section 5).

The semigroups $\E^+$ and $\E_{\ast}$
have been studied in
\cite{ES3}.
We proved in \cite{ES3} a certain finiteness
property for subsets of the somewhat smaller
$\E_{\ast}$ which behave like sets of conformal
characters. More precisely, we call a finite subset $S$ of
$\Z_{>0}\cdot\E^+$ modular
if its additive span (i.e.\ the subspace of the space of all holomorphic
functions on the upper half plane which is spanned by the functions in $S$)
is invariant under $\SL$.
To study modular subsets of $\E_{\ast}$ one can use
a natural sort of graded filtration on this semigroup.
Namely, it is easy to show that each modular subset of $\E_{\ast}$
is actually contained in some $\E_n(l)$, the set of all modular 
units in $\E_{\ast}$ consisting of at most $n$-fold products
of the functions $[1]_l, \dots, [\lfloor l/2\rfloor ]_l$.
Then the finiteness property says that, for a given $n$, there exist
only a finite number of $l$ such that $\E_n(l)$
contains a modular subset.

For $n=1,2,3$ a complete list of all modular subsets of $\E_n(l)$ has 
been given loc.\ cit.. Surprisingly, all these
and any other modular set which is known to us, are sets
of conformal characters of rational models
of $\w$-algebras.

For brevity we call a RCFT modular
if its characters
form a modular set.
It follows from the results of \cite{ES3} that the effective
central charge of a modular RCFT whose characters
are in $\E_n(l)$ is given by
$\tilde c = \frac{2n}{l}$.

Some examples of modular RCFTs have been known for quite some time: 
The characters of the Virasoro minimal models with central
charge $c = c(2,q)$ and $c=c(3,q)$ have product 
expansions (see e.g.\ \cite[Section 2]{KRV} and ref. therein), 
and can thereby be identified as modular units. These examples also play  
an interesting role in another context:
they are equal to one side of the famous generalized Rogers-Ramanujan
identities (see e.g.\ \cite{A} for a review on their occurrence 
in various subjects). 
The other side of these identities has a certain sum form, called 
fermionic in the physical literature, which is believed to encode 
information about integrable perturbations of the RCFT
under consideration (see e.g.\ \cite{KKMM}). 
It has even been conjectured that any character of a 
RCFT has such a fermionic sum formula 
(independently of being a modular unit or not). 
For a proof of this conjecture in the case of the Virasoro minimal 
models and more details see \cite{BM}.

The plan of the rest of this letter is as follows:
In Section 3 we explain four infinite series of modular RCFTs
with effective central charge $\tilde c\le1$.
For two of these series (3.1 and 3.4) the product formulas are
classical facts in the mathematical literature, for
the two others they seem to be new; in any case we explain the relevant tools
to prove the product formulas.
In Section 4 we give examples of modular RCFTs with $\tilde c>1$.
Furthermore, we conjecture 
that certain infinite series of minimal models of 
Casimir $\w$-algebras are modular.
Finally, in Section 5 we pose and discuss some open questions
which naturally arise in our investigations.  
In Section 2 we describe an
algorithm for testing whether a given conformal 
character is a modular unit or not.
The examples in Section 3.3 and Section 4 have
been found using this algorithm.

\section{A modular unit test}

In this section we describe an algorithm which can be used 
for testing whether a given conformal character is a modular unit. 
The algorithm can be easily implemented as a computer program.
It is based on two observations, which we shall now explain.

The first observation is that
any periodic and holomorphic function on the upper half plane
has a unique product expansion, which can easily be calculated. 
More precisely one has 

{\it 
Let $f$ be a holomorphic function on the upper half plane,
assume that $f(z+1)=f(z)$ (so that $f(z)$ can be written as power
series in $q=\e^{2\pi iz}$), and that the Fourier expansion of $f$
is in $1+q\,\C[\![q]\!]$.
Then there exists a unique sequence $a(n)$ of complex numbers
such that
$$
f(z)=\prod_{n\ge1}(1-q^n)^{a(n)},
$$
for sufficiently small $|q|$.
(Here, for complex $\alpha$, we use
$
(1-q^n)^{\alpha}=\exp\left(-\alpha\sum_{k\ge 1}\frac{q^{nk}}k\right)
$.)
}

Note that it  is easy to prove that the $a(n)$ are actually 
integers if $f$ has integral Fourier coefficients \cite[Lemma 3]{ES3}.

The proof of this statement is a consequence of the following simple 
computation
\begin{multline*}
q\frac d{dq}\log f
=
\sum_{l\ge1} c(l)\,q^l
=
\sum_{l\ge1}  \big(\sum_{n|l}b(n)\big)q^l \\
=
\sum_{n,k\ge1}b(n)\,q^{nk}
=
\sum_{n\ge1}\frac{b(n)\,q^n}{1-q^n}
= 
q\frac d{dq}
\log \prod_{n\ge1}(1-q^n)^{-b(n)/n},
\end{multline*}
where
$b(n)=\sum_{d|n}c(d)\mu(n/d)$ with the
the Moebius function $\mu(d)$, i.e. with
the arithmetic function defined by
$\mu(d)=(-1)^\nu$ ($\nu=$ number of prime factors of $d$), if $d$
is squarefree, and $\mu(d)=0$ otherwise.

The second observation is as follows.
Let $S$ be the set of conformal characters of a RCFT. Assume
that the conformal characters in $S$ are modular functions
on a congruence subgroup of $\SL$ (as is true in all known
examples, and as is believed to hold true in general). Let
$l'$ be the smallest positive integer such that $\chi(z+l')=\chi(z)$,
for all $\chi\in S$, or, equivalently, let $l'$ be the denominator of the numbers $h-c/24$,
where $c$ is the central charge and where $h$ runs through the set
of conformal dimensions.
Then each $\chi$ is a modular function on
the principal congruence subgroup $\Gamma(l')$ \cite[Section 4.3]{ES1}.
On the other hand, a modular function $\chi$ on $\Gamma(l')$ with
Fourier expansion of the form $q^s\Z[\![q]\!]$ is a
modular unit if and only if
$$
\tilde\chi=\prod_{1\le j \le \lfloor l/2\rfloor}[j]_l^{a(j)}
$$
for suitable integers $a(j)$ \cite[Theorem 1]{ES3}.
Here $l=l'$, if $l'$ is odd, and $l=2l'$, if $l'$
is even, and $\tilde\chi=\chi/A$ where $A$ is the first nonzero
Fourier coefficient of $\chi$.

Note that to verify, for given integers $a(j)$, that
$$f:=\tilde\chi-\prod_{1\le j \le \lfloor l/2\rfloor}[j]_l^{a(j)}$$
is identically zero,
it suffices to check 
$f\equiv 0\bmod q^N$
for a sufficiently large $N$.
Namely,
$[r]_l$ is a modular function on $\Gamma(12l)$, with
orders in the cusps between $-1/12l$ and $l/24$
\cite[Section 2]{ES3}. The orders in the cusps of $\tilde\chi$
are always $\ge -\tilde c/24$, where $\tilde c$ is the effective central charge
of the corresponding RCFT. Thus, $f$ is a modular function
on $\Gamma(12l)$ whose pole order $t_s$ in an arbitrary cusp $s$
is bounded from below by
$$
t_s\ge t:=-\frac{\tilde c}{24}
-\frac1{12l} \sum_{a(j)>0}a(j)
+\frac l{24} \sum_{a(j)<0}a(j),
$$
and without any pole in the upper half plane.
Since a nonzero modular function
cannot have more zeros than poles
(counted with multiplicities in suitable local parameters
on the compactified Riemann surface associated to $\Gamma(12l)$),
we see that it suffices to check the above congruence
for
$$
N>-t\cdot \big(-1+\frac12(12l)^2\prod_{\pi|6l}(1-\frac1{\pi^2})\big),
$$
(The second factor on the right is the number of cusps of $\Gamma(12l)$
minus 1, the product has to be taken over all primes $\pi$ dividing $6l$.)

The algorithm for testing whether a function $\chi\in S$ 
is a modular unit can now be described  as follows
(recall that $\tilde\chi=\chi/A$ with $A$ as the first nonzero
Fourier coefficient of $\chi$):
\begin{list}{}{}
\item[{\bf Step 1:}]
  Compute the first $\lfloor l/2\rfloor$ many complex numbers $a(j)$, accordingly to the statement above, so that 
  $$q^{h-c/24}\tilde\chi\equiv
  \prod_{1\le j \le \lfloor l/2\rfloor}(1-q^j)^{a(j)}
  \bmod q^{1+\lfloor l/2\rfloor}.$$
  If all these $a(j)$ are integers then do Step 2. Otherwise
  $\chi$ is not a modular unit.
\item[{\bf Step 2:}]
  Check if
  $$\tilde\chi\equiv \prod_{1\le j\le\lfloor l/2\rfloor}[j]_l^{a(j)}\bmod  q^N,$$
  where $N$ is as above.
  If the answer is positive, then we actually have equality. Otherwise,
  $\chi$ is not a modular unit.
\end{list}

In practice it turns out that the conformal characters can
often be written as products of $[r]_{l'}$ with $l'$ a divisor of $l$.
For passing from one product representation to the other one uses the
easy distribution property
$$
[r]_{l'}
=\prod_{
\begin{subarray}{c}
s\bmod l\\ s\equiv r\bmod l'
\end{subarray}}
[s]_l.
$$

\section{Modular RCFTs with $\tilde c = \frac{2n}{l}\le 1$}

We give four infinite series of modular RCFTs, two of which are series of
Virasoro minimal models, and the third one closely related.
The fourth series is given by certain extensions of the current 
algebra $\widehat{U(1)}$.

Let us recall first some basic facts about Virasoro minimal models.
These models have pairwise different central charges. The set of
their central charges equals the set of all numbers
$c(p,q) = 1 - 6\frac{(p-q)^2}{pq}$, where $p$ and $q$ are coprime
integers greater than 1.
Note that $c(p,q)=c(p',q')$ implies
$\{p,q\}=\{p',q'\}$.
The effective central charge of the model
with $c=c(p,q)$ is given by $\tilde c(p,q) = 1 -6\frac1{pq}$.
If we take $p$ odd, then its (pairwise different) conformal characters are given by
$$
\chi^{p,q}_{r,s} = \frac{1}{\eta} 
               \left( \theta_{rp-sq,pq}-\theta_{rp+sq,pq} \right)
\qquad (r=1,\dots,q-1; s=1,\dots,(p-1)/2),
$$
where
$$
\theta_{\lambda,k}
=\sum_{\begin{subarray}{c}n\in\Z\\ n\equiv \lambda \bmod 2k\end{subarray}}
q^{n^2/4k},\qquad
\eta=q^{1/24}\prod_{n\ge 1}(1-q^n).
$$
(The reader may excuse the double usage of $q$ in these formulas,
which is due to an interference of traditional notations).
These characters are identical to the characters of
certain irreducible representations   
of the Virasoro algebra.
It is well-known that the space spanned by these
characters is invariant under $\SL$ (see e.g.\ \cite{S}).

\subsection{The Virasoro minimal models $c=c(l,2)$}

The set $S$ of characters 
of the Virasoro minimal model $c(l,2)$ is modular. The characters
read
$$ \chi^{l,2}_{1,s}
=
\pm\,\frac{1}\eta
\sum_{n\equiv 2s\bmod l}
\left(\frac{-4}n\right)\,q^{n^2/8l}
=
\prod
_{\begin{subarray}{c}1\le j\le\frac{l-1}2\\ j\not=s\end{subarray}}
[j]_l
\qquad(1\le s\le\textstyle\frac{l-1}2).
$$
Here and in the following, for an integer $D\equiv 0,1\bmod 4$, we use
$\left(\frac D \cdot\right)$ for the unique Dirichlet character modulo $D$
such that, for any odd prime $n$ not dividing $D$, one has 
$\left(\frac D n\right)=1,-1$ accordingly as $D$ is a 
square modulo $n$ or not, and such that $\left(\frac D n\right)=0$
if $\gcd(n,D)\not=1$.
The product expansions follow on writing in the above sum formula
$n=2s+l+2lm$
(so that $\left(\frac{-4}n\right)=\left(\frac{-4}{2s+l}\right)(-1)^m$)
and on applying the Jacobi 
triple product identity \cite[p.\ 282]{H-W}
$$
\sum_{m\in\Z} u^{m^2}v^m=\prod_{n\ge1}(1-u^{2n})(1-q^{2n-1}v)(1-q^{2n-1}/v)
$$
with $u=q^{l/2}$ and $v=-q^{s+l/2}$.
(One also needs that the first factor in the Jacobi 
triple product identity becomes $\eta(lz)$, and that $\eta(lz)/\eta(z)=[1]_l[2]_l\cdots[(l-1)/2]_l$.)
The product expansions of the $\chi^{l,2}_{1,s}$ have already been noted in
\cite[p.\ 137]{S} and, in the context of Virasoro minimal models, in 
\cite{C}.

\subsection{The Virasoro minimal models $c=c(3,l)$}

Again the set $S$ 
of characters of the Virasoro minimal model $c(3,l)$ is modular.
The characters are
$$ \chi^{3,l}_{l-r,1}=
\pm\,\frac{1}\eta
\sum_{n\equiv 3r\bmod 2l}
\left(\frac{-3}n\right)\,q^{n^2/12l}
=
\prod
_{\begin{subarray}{c}
1\le j\le 2l-1\\
j\not=2r,l-r,l+r
\end{subarray}}
[j]_{4l},
$$
where
$1\le r\le l-1$.
To obtain the product expansions one splits the sum over $n$ into two sums over
$n=3r+2l+6lm$ and $n=3r-2l-6lm$, respectively. The Dirichlet character is then identically
equal to $+1$
in one, and to $-1$ in the other sum. Thus
$$
\chi^{3,l}_{l-r,1}
=
\pm\,\frac{q^{(4l^2+9r^2)/12l}}\eta
\sum_{m\in\Z}
\left(q^{l(3m+2)m+(3m+1)r}-q^{l(3m+2)m-(3m+1)r}\right).
$$
Now we apply the quintuple product formula \cite[p.\ 185]{Kac}
\begin{multline*}
\sum_{m\in\Z}
u^{(3m+2)m}
\left(v^{3m+1}-v^{-3m-1}\right)
=
(v-v^{-1})\cdot\\
\cdot\prod_{n\ge 1}
(1-u^{2n})(1-u^{2n-1}v)(1-u^{2n-1}/v)(1-u^{4n}v^2)(1-u^{4n}/v^2)
\end{multline*}
(Note that loc.cit.\ one has to substitute $(u,v)\leftarrow (u/v,v^2)$
to obtain the formula as stated here.)
Setting $u=q^{l}$ and $v=q^{r}$ gives
$$
\chi^{3,l}_{l-r,1}=\frac{\eta(2lz)}{\eta(z)\,[r+l]_{2l}\,[2r]_{4l}}.
$$
Rewriting this in
terms of functions $[j]_{4l}$
on using 
the distribution property $[j]_{2l}=[j]_{4l}\,[j+2l]_{4l}$ 
and on expanding $\eta(2lz)/\eta(z)$ as product of $[j]_{4l}$'s,
one obtains the stated product expansions of the characters.
(Warning: if one tries to expand $\eta(2lz)/\eta(z)$ as a product 
of $[j]_{2l}$'s, then a fractional power $[l]_{2l}^{1/2}$ will occur;
here one has to use the simple identity $[l]_{2l}^{1/2}=[l]_{4l}$.) 
These product expansions have been similarly obtained in 
\cite[Eq. 2.9]{KRV} (as special cases of the 
MacDonald identity for the Cartan matrix $(2,-4;-1,2)$). 

The first nontrivial example $l=4$ of this series ($l=2$ is the trivial
Virasoro minimal model with character 1) is the Ising model with central
charge $\tilde c(3,4) = c(3,4)=1/2$, which gives
$$S 
=
\{ [2, 3, 4, 5]_{16}, [1, 4, 6, 7]_{16}, [1, 3, 5, 7]_{16} \}
\subset \E_4(16). $$ 
Here we have used   
$[a_1,\dots,a_n]_l$ for $\prod_{i=1}^n [a_i]_l$.

\subsection{$\w$-algebras of type $\w(2,\frac{l-1}{2})$
with $c = c(2l,3)$}

Let
$l>1$ and relatively prime to 6.
For an odd integer $1\le r\le l$ let
$$\chi_r =
	   \sigma_r\cdot 
           \bigl( \chi^{2l,3}_{1,r} + 
                  \chi^{2l,3}_{1,2l-r}\bigr)
=
\pm\,\frac{1}\eta
\sum_{n\equiv 3r\bmod l}
\left(\frac{12}n\right)\,q^{n^2/24l}
$$
($\sigma_l=\frac12$ and $\sigma_r=1$ for $r\not=l$).
These functions have been identified as the conformal characters 
of rational models of certain $\w$-algebras. More precisely
they are the characters of $\w$-algebras of type $\w(2,\frac{l-1}{2})$
with central charge $c = c(2l,3)$ \cite[Section 4]{EFHHNV}.

The transformation properties 
of the theta function $\theta_{\lambda,k}$ \cite[p.\ 37]{S} 
show that the space spanned  by the functions $\chi_r$ ($r=1,3,\dots,l$) is 
invariant under $\SL$.

The $\chi_r$ are also modular units. If $r=l$, then
$\chi_r(z)=\eta(lz)/\eta(z)$, whence
$$
\chi_l=\prod_{1\le r\le \frac{l-1}2}[r]_l.
$$

The product expansions for the $\chi_r$ with $r<l$ are more difficult.
They can be proved as follows:
The coefficient of
$q^{n^2/24l}$ in the above sum formula for $\chi_r$
is different from 0 only if
$n\equiv 3r\bmod l$ and $n\equiv \pm1\bmod 6$.
We decompose the sum into two sums accordingly as $n$ is congruent
to $+1$ or $-1$ modulo 6.
We can write these as one sum over
$n=2l+3r+6lm$ and a second one over $n=-2l+3r-6lm$.
This is possible since $l$ is not divisible by 3 and since $r$ is odd.
The Dirichlet character equals $\left(\frac{12}{2l+3r}\right)(-1)^m$
in the first, and it equals
$\left(\frac{12}{-2l+3r}\right)(-1)^m=-\left(\frac{12}{2l+3r}\right)(-1)^m$
in the second. For the last identity we needed $2l\equiv\pm2\bmod12$.
We thus obtain
$$
\chi_r
=
\pm\,\frac{q^{(4l^2+9r^2)/24l}}{\eta}
\sum_{m\in\Z}
(-1)^m\Big(
q^{\frac{l(3m+2)m}2}q^{\frac{3rm+r}2}
-
q^{\frac{l(3m+2)m}2}q^{-\frac{3rm+r}2}
\Big).
$$
Now we apply the quintuple product formula
with $u=q^{l/2}$ and $v=-q^{r/2}$.
This gives the product expansion
$$
\chi_r
=
\frac{\eta(lz)\,[(r+l)/2]_l}{\eta(z)\,[r+l]_{2l}\,[r]_{2l}}.
$$
(Here we rewrote factors $(1+q^s)$ as $(1-q^{2s})/(1-q^s)$.)
Using the distribution property
$[r+l]_{2l}[r]_{2l}=[r]_l$ this
can be rewritten as
$$
\chi_r=[(r+l)/2]_{l}
\prod_{
\begin{subarray}{c}
1\le j \le \frac{l-1}2\\ j\not\equiv\pm r\bmod l
\end{subarray}}
[j]_{l}
$$
Note that here, in contrast to the preceding examples,
certain $[j]_{l}$ occur with multiplicity 2
in the product expansions of the $\chi_r$.
The first non-trivial case is $l=7$, i.e.\ the
rational model of $\w(2,3)$ with $c=c(14,3)=-\frac{114}{7}$. Here one finds 
$$ S = \{ [2,3,3]_7, [1,2,2]_7, [1,1,3]_7, [1,2,3]_7 \} \subset \E_3(7). $$
This is the modular set $W_7$ which was found in \cite{ES3}.

\subsection{The extended $\widehat{U(1)}$ theory}

In this subsection we give an example of a modular set which 
is not contained in $\E_{\ast}$.
 
Consider the conformal field theory given by a single $\widehat{U(1)}$ 
current living on a circle of radius $R$. This conformal 
field theory has central charge $c = 1$ and is not rational. 
However, if $l := 2 R^2$ is an integer this theory can be extended 
by two primary fields of conformal dimension $l$ and then becomes 
rational (see e.g. \cite[Section 5]{MS}). The effective central charge 
of this RCFT equals the central charge $\tilde c = c =1$.   
The conformal characters of its finitely many inequivalent 
irreducible representations are given by 
$ \chi_m = \theta_{m,l}/\eta$ ($0\le m<2l$).
As is well-known, these functions span a $\SL$-module

Using the Jacobi triple product identity \cite[p.\ 282]{H-W}
one finds that 
$$  
  \chi_0 = \frac{[l]_{4l}^2}{[2l]_{4l}} \prod_{1\le j <2l} [j]_{4l}, \quad
  \chi_l =  [2l]_{4l}     \prod_{\begin{subarray}{c} 
                                  1\le j<2l
 				  \end{subarray}} 
                                [j]_{4l},\quad
  \chi_m = [m-l]_{4l} [m+l]_{4l}
                           \prod_{\begin{subarray}{c} 
                                  1\le j<2l \\
                                  j \not\equiv \pm 2(m-l) \bmod 4l
                                  \end{subarray}} 
                                [j]_{4l}
$$
where $1\le m<2l$ and  $m\not=l$.

Therefore, the set of the conformal characters $\chi_m$ ($0\le m<2l$) 
is modular.
Note that the vacuum character $\chi_0$, being an element of $E^+$,
is not contained in $E_*$, since it contains a negative power
of $[2l]_{4l}$.

\section{Further examples and (conjectured) infinite series}

In this section we present three more subtle examples
of rational models whose conformal characters all are 
modular units. We give the product expansions explicitly 
and show that the corresponding sets $S$ are modular.
We conjecture that these cases are only the first examples 
contained in infinite series of rational models of Casimir 
$\w$-algebras whose set of conformal characters is modular,
i.e.\ where all characters have `nice' product expansions.
The precise form of the conjecture is given below. 

Let us first discuss the three examples where we can prove
that they are modular RCFTs. These three rational models of
$\w$-algebras were
considered in \cite{ES1,ES2}, namely
$\w(2,4)$ at $c=-\frac{ 444}{11}$, 
$\w(2,6)$ at $c=-\frac{1420}{17}$ and 
$\w(2,8)$ at $c=-\frac{3164}{23}$.
Note that $\w$-algebras of type
$\w(2,4)$ and $\w(2,6)$ are Casimir $\w$-algebras associated
to the simple Lie-algebras ${\cal B}_2$ and ${\cal G}_2$, respectively,
and that 
$\w(2,8)$ at $c=-\frac{3164}{23}$ is a truncation of the Casimir 
$\w$-algebra ${\cal WE}_7$ \cite{EHH}.
The sets of conformal dimensions of these rational models can e.g.\ be 
found in \cite[Table 2, p.\ 125]{ES1}. 
General expressions for conformal characters of rational models 
of Casimir $\w$-algebras and their transformation properties 
have been obtained in \cite[Proposition 3.4(b), p.\ 320]{FKW}.
In the case of the three rational models under consideration 
formulas have been independently worked out in \cite{ES2} 
using theta series associated to quaternion algebras. 
These latter formulas allow to compute conveniently
the conformal characters in question up to any order.
One can therefore use the algorithm described in section 2
to prove that all the characters of the three rational models  
are modular units and that their product expansions are those given 
in the table below. 
$$
\begin{array}{|l|c|c||l|l|}
%\multicolumn{4}{c}{\strut\text{Table: Conformal characters as modular units}}\\
    \hline
    \text{$\w$-alg.}&c&\tilde c&\text{conformal characters}&\text{in}\\
    \hline
    \w(2,4)&-\frac{444}{11}&\frac{12}{11}&[1,1,2,3,4,5]_{11},&\E_6(11)\\ 
           &&&[2,3,4,4,5,5]_{11}&\\
    \hline
    \w(2,6)&-\frac{1420}{17}&\frac{20}{17}&[1,1,2,3,4,5,5,6,7,8]_{17},&\E_{10}(17)\\ 
           &&&[2,3,4,5,6,6,7,7,8,8]_{17}&\\
    \hline
    \w(2,8)&-\frac{3164}{23}&\frac{28}{23}&[1,1,2,3,4,5,5,6,7,7,\ 8,\ 9,10,11]_{23},&\E_{14}(23)\\
           &&&[2,3,4,5,6,7,8,8,9,9,10,10,11,11]_{23}&\\
    \hline
   \end{array}
$$
Here, for each $\w$-algebra,  we listed only two product expansions 
$[r_1,\dots,r_n]_l=\prod_{j=1}^n[r_j]_l$; one has to add all products of the form 
$[s\cdot r_1,\dots,s\cdot r_n]_l$ with $2\le s\le \frac{l-1}2$, respectively.

We conjecture that these modular sets are only the first few members of
infinite series of rational models whose conformal characters are
modular units. In order to be more precise we need to fix some notations.

Let ${\cal L}_k$ be a finite-dimensional simple Lie algebra of rank $k$
over $\C$, let  $\rho$ ($\rho^\vee$) the sum
of its (dual) fundamental weights, $h$ ($h^\vee$) its (dual) Coxeter
number, $e_r$ ($1\le r\le k$) its exponents, and denote
by ${\cal WL}_k$ the one-parameter family (parametrized by the central charge)
of corresponding Casimir $\w$-algebras \cite[pp.\ 207]{BS}.
One expects that the minimal models of ${\cal WL}_k$, i.e.\ the rational models
of ${\cal WL}_k$, are those where the central charge $c$ can be written in the form
$
c=c(p,q):=k-12(q\rho-p\rho^\vee)^2/pq
$
with integers $p,q$ satisfying
$p\ge h^\vee, q\ge h$ and $\gcd(p,q) = 1$.
Such a pair $p,q$ is called a parametrization of $c$.
It is called minimal if
$p\cdot q\le p'\cdot q'$ for all other parametrization $p',q'$
of $c$.
\begin{conj}
Let $S^{{\cal L}_k}_l$ be the set of conformal characters of
the rational model of the Casimir $\w$-algebra
${\cal WL}_k$ with central charge $c = c^{{\cal L}_k}(h^\vee,l)$.
Assume that $h^\vee,l$ is a minimal parametri\-zation.
Then one has
\begin{list}{}{}
\item[(1)]
   $S^{{\cal L}_k}_l$ is modular.
   The product expansions of the characters $\chi\in S$
   are of the form
   $ \chi = \prod_{j=1}^{\lfloor l/2\rfloor} [j]_l^{a(j)} $
   with non-negative integers $a(j)$.
\item[(2)]  
   If $l\ge 2h$ the vacuum character is given by    
   $$
	\chi_0 =
	\begin{cases}
		\phantom{[l/2]_l^{k/2}}
		\prod_{i=1}^{k} \prod_{e_i<j<l/2} [j]_l
			&\text{for $l$ odd}\cr
		[l/2]_{2l}^k\prod_{i=1}^{k} \prod_{e_i<j<l/2} [j]_l
			&\text{for $l$ even.}
	\end{cases}
   $$
\end{list}
\end{conj}
Thus, for odd $l$, the $a(j) $ equal the number of exponents of
${\cal L}_k$ smaller than $j$. 
Note that the condition $l\ge 2h$ in $(2)$ ensures that
each product over $e_i<j\le l/2$ really contributes to the 
character ($\max_r e_r = h-1$). For $l<2h$ one might expect 
that a truncation of the Casimir $\w$-algebra in question occurs. 
An example of this phenomenon is given by the third rational model 
discussed at the beginning of the section. For questions related 
to truncations of Casimir $\w$-algebras see \cite{BEHHH}. 

We explain why we believe this conjecture to hold true.

Firstly, the conjecture is true for 
${\cal L}_k = {\cal A}_1$. In this case the minimal models of ${\cal WL}_k$
are the Virasoro minimal models discussed in Section 3.1.

Secondly, the conjecture is true for the three rational models
discussed at the beginning of this section.

Thirdly, we have used the formulas of \cite[p.\ 320]{FKW}
and a computer
program written in PARI-GP \cite{GP} to test our conjecture for 
${\cal A}_2$, ${\cal B}_2$ and ${\cal G}_2$ and low values of $l$.

Finally,
we have two theoretical arguments for our conjecture.
The first one is the following: From the explicit form of 
the vacuum character for $l\ge 2h$ it is easy to calculate 
the central charge, i.e.\ the vanishing order of 
$\chi_0$ at $i\infty$ multiplied by $-24$.
This value agrees with the value given 
by the formula stated in the conjecture.

The second argument relies on the fact that one can calculate 
the first Fourier coefficients of the vacuum character by 
looking at the Kac determinants of the vacuum representation.
 
More precisely, one can show that the vacuum character is generic
up to order $N-1$, i.e.\
$$\chi =  \prod_{i=1}^k \prod_{j>e_i}  (1-q^j)^{-1}  + \LO(q^N),$$  
if and only if the factors of the Kac determinants for all levels $<N$ 
either do not vanish or vanish generically. 
The maximal $N$ for which the above equation holds true
equals the minimal level where a generically non-vanishing factor of 
the Kac determinant at this level is zero.
It has been observed in \cite[p.\ 58]{BEHHH} that, for a minimal 
parametrization $p=h^\vee-1+r,p'=h-1+s$ of the central charge 
$c = c^{{\cal L}_k}(p,p')$, this level is given by $N=rs$. 
In our case we have $r=1$ and $s=l+1-h$ so that $N=l+1-h$.  
Using the well-known fact that, for a simple Lie algebra, 
$\max_r e_r = h-1$ we find agreement with our conjectured form of 
the vacuum character for $l \ge 2h$.

\section{Conclusions}

We have described a simple 
algorithm which allows to determine whether
a given conformal character (which is known to be
a modular function on some congruence subgroup)
is a modular unit or not. Using this 
algorithm we have found highly non-trivial examples of modular
sets, sets of certain modular 
units which span $\SL$-modules. 
All known examples of such modular sets are related to RCFTs. 

There are several interesting open questions. 

The most obvious question is whether all modular sets are related to 
RCFTs. This question seems to be very difficult to answer since neither 
a classification of modular sets nor a classification of RCFTs does 
exist so far. However, it might be possible to answer the question 
at least for those cases where $2n<l$, i.e.\ where the effective 
central charge of the corresponding RCFTs is smaller than one.
In this case more information, on the side of $\w$-algebras, as well
as on the side of modular forms, is available. 
We intend to study this in a future publication.

Furthermore, it would be very interesting if the new examples presented
in this letter could be  interpreted as parts of new generalized 
Rogers-Ramanujan type identities.

Finally, the combinatoric form of the conjectured product expansions 
of the conformal characters of minimal models of Casimir $\w$-algebras
${\cal WL}_k$ with central charge $c = c^{{\cal L}_k}(h^\vee,l)$ seems 
to suggests that there exist very special bases of the irreducible 
representations. Looking for example at the vacuum character 
$ \chi_0 = [2]_5$
%$$ = \prod_{n\equiv \pm 2 \bmod 5} (1-q^n)^{-1} $$
of the Virasoro minimal model with central charge $c=-22/5$ one 
might guess that already 
the states of the form 
$ L_{-n_k}\dots L_{-n_1}\nu_0$ with
$n_k\ge \dots\ge n_1$ and $n_i \equiv \pm 2 \bmod 5$
form a basis of the (irreducible) vacuum representation.

%%%%%%%%%%%%%%%%%%%%%%
%% Acknowledgements %%
%%%%%%%%%%%%%%%%%%%%%%
\bigskip
{\bf Acknowledgments.}
We would like to thank  M.\ R\"osgen and G.\ Tak\'acs
for discussions and A.\ Honecker, H.\ Kausch and M.\ Gaberdiel 
for comments on a draft version of this letter.
N.-P.\ Skoruppa thanks the Department for Applied Mathematics
and Theoretical Physics for financial support during his stay in 
Cambridge.


\small
\begin{thebibliography}{10}
\bibitem[1]{A}
G.E.\ Andrews,
{\it $q$-Series: Their Development and Application in Analysis, Nuber
Theory, Combinatorics, Physics, and Computer Algebra},
Conference Board of the Mathematical Sciences, Regional Conference
Series, Number 66, American Mathematical Society (1986).

\bibitem[2]{GP}
C.\ Batut, D.\ Bernardi, H.\ Cohen, M.\ Olivier,
{\it PARI-GP}, Universit\'e Bordeaux 1, Bordeaux (1989).

\bibitem[3]{BM}
A.\ Berkovich, B.M.\ McCoy,
Lett. Math. Phys. 37 (1996), 49-66.

A.\ Berkovich, B.M.\ McCoy, A.\ Schilling,
preprint q-alg/9607020,
Commun. Math. Phys. (to appear).

\bibitem[4]{BEHHH}
R.\ Blumenhagen, W.\ Eholzer, A.\ Honecker, K.\ Hornfeck, R.\ H{\"u}bel,
Phys. Lett. B332 (1994), 51-60.

\bibitem[5]{BS} 
P.\ Bouwknegt, K.\ Schoutens,
Phys. Rep. 223 (1993), 183-276.

\bibitem[6]{C} 
P.\ Christe,
Int. J. Mod. Phys. A6 (1991), 5271-5286.

\bibitem[7]{EFHHNV}
W.\ Eholzer, M.\ Flohr, A.\ Honecker, R.\ H{\"u}bel, W.\ Nahm, 
R.\ Varnhagen,  
Nucl. Phys. B383 (1992), 249-288.

\bibitem[8]{EHH}
W. Eholzer, A. Honecker, R. H{\"u}bel,
Phys. Lett. B308 (1993), 42-50.

\bibitem[9]{ES1}
W.\ Eholzer, N.-P.\ Skoruppa,
Commun. Math. Phys. 174 (1995), 117-136.

\bibitem[10]{ES2}
W.\ Eholzer, N.-P. Skoruppa,
Lett. Math. Phys. 35 (1995), 197-211.

\bibitem[11]{ES3}
W.\ Eholzer, N.-P.\ Skoruppa,
%{\it $\SL$ --- invariant spaces spanned by modular units},
preprint DAMTP-96-62.
%, q-alg/96mmnnn.

\bibitem[12]{FKW}
E.\ Frenkel, V.\ Kac, M.\ Wakimoto,
Commun. Math. Phys. 147 (1992), 295-328.

\bibitem[13]{H} 
J.E.\ Humphreys, 
{\it Introduction to Lie Algebras and Representation Theory}, 
Springer-Verlag, New York, Heidelberg, Berlin (1972).

\bibitem[14]{H-W}
G.H.\ Hardy, E.M.\ Wright,
{\it An introduction to the theory of numbers},
4th edition,
Oxford University Press, Oxford (1975).

\bibitem[15]{Kac}
V.G.\ Kac,
{\it Infinite Dimensional Lie Algebras},
3rd edition,
Cambridge University Press, Cambridge (1990).

\bibitem[16]{KKMM}
R.\ Kedem, T.R.\ Klassen, B.M.\ McCoy, E.\ Melzer,
Phys. Lett. B304 (1993), 263-270.

\bibitem[17]{KRV}
J.\ Kellendonk, M.\ R\"osgen, R.\ Varnhagen,
Int. J. Mod. Phys. A9 (1994), 1009-1024.

\bibitem[18]{MS}
G.\ Moore, N.\ Seiberg,
Nucl. Phys. B313 (1989), 16-40. 

\bibitem[19]{S}
N-P.\ Skoruppa,
{\it \"Uber den Zusammenhang zwischen
Jacobiformen und Modulformen halbganzen Gewichts},
Bonner Mathematische Schriften Nr.\ 159,
Bonn (1985).

\end{thebibliography}
\end{document}